\documentclass[twocolumn,showpacs,preprintnumbers,amsmath,amssymb,superscriptaddress]{revtex4}

\usepackage{graphicx}
\usepackage{dcolumn}
\usepackage{bm}
\usepackage{tabularx}
\usepackage{ulem} 
\newcolumntype{M}{>{\centering\arraybackslash}m{1.85cm}}
\usepackage[export]{adjustbox}
\usepackage{float}
\usepackage[caption=false]{subfig}
\usepackage{color}   
\usepackage{hyperref}
\hypersetup{
    colorlinks=true, 
    linktoc=all,     
    linkcolor=blue,  
}

\makeatletter
\newcommand{\colorcaption}[2][]{%
  \begingroup%
  \renewcommand{\@caption@fignum@sep}{ (Color online). }%
  \caption[#1]{#2}%
  \endgroup%
}
\makeatother
\newcommand\T{\rule{0pt}{3ex}}       
\newcommand\B{\rule[-1.5ex]{0pt}{0pt}} 

\begin{document}

\title{Structure of $^{46,47}$Ca from the $\beta^-$ decay of $^{46,47}$K 
in the framework of nuclear shell model}

\author{ Priyanka Choudhary}
\address{Department of Physics, Indian Institute of Technology Roorkee, Roorkee 247667, India}
\author{Anil Kumar}
\address{Department of Physics, Indian Institute of Technology Roorkee, Roorkee 247667, India}
\author{Praveen C. Srivastava}
\email{praveen.srivastava@ph.iitr.ac.in}
\address{Department of Physics, Indian Institute of Technology Roorkee, Roorkee 247667, India}
\author{Toshio Suzuki}
\address{Department of Physics, College of Humanities and Sciences, Nihon University, Sakurajosui 3, Setagaya-ku, Tokyo 156-8550, Japan}

\date{\hfill \today}
\begin{abstract}

In the present work, we report a comprehensive theoretical study of the $\log ft$ values for the allowed and forbidden $\beta^-$ decay transitions of $^{46}$K and $^{47}$K  corresponding to recently available experimental data from the GRIFFIN spectrometer at TRIUMF-ISAC [Phys. Rev. C {\bf 100}, 054327 (2019); Phys. Rev. C {\bf 102}, 054314 (2020)]. We perform the nuclear shell-model calculation in $sdpf$-valence space, with the SDPF-MU interaction, to calculate low-lying energy spectra of $^{46,47}$K and $^{46,47}$Ca. For further investigation, we also calculate spectroscopic properties of $^{46,47}$K, $^{46,47}$Ca and compare the results with the experimental data. We suggest spin-parity of several levels, which were previously tentative, in both the calcium isotopes. Based on the energy and $\log ft$ values, we conclude that the state at 3.984 MeV in $^{46}$Ca could be either $3^{-}$ or $2^{+}$ and a state at 4.432 MeV might be $3^{-}$. The spin-parity of the states at 2.875, 3.951 and 4.453 MeV are predicted to be $1/2^-$, $3/2^-$ and $1/2^-$ or $1/2^+$, respectively, in $^{47}$Ca. We also give the confirmation to spin-parity of 2.850, 3.889 and 4.606 MeV states for $^{47}$Ca. 
 Our results of level schemes of $^{46,47}$K are in a reasonable agreement with the experimental data, 
while the agreement of calculated $\log ft$ values for the $\beta^-$ decay with the experimental data is limited.
 
\end{abstract}

\maketitle
\section{Introduction}
Understanding the formation of shell structure across an isotopic chain (from proton drip line to neutron drip line) is a key challenge in nuclear physics.
New experiments have revealed the appearance of new shell closures N = 16, 32, 34 \cite{A.Gade, F.Wienholtz, D.Steppenbeck, D.SteppenbeckPRL, M.Rosenbusch, H.N.Liu, A.Ozawa,R.Kanungo} and the disappearance of the magic numbers N = 8, 20, 28 \cite{A.Navin,T.Motobayashi, B.Bastin}.
Evolution of shell structure in calcium isotopes has been a crucial area of investigation in nuclear structure over the past few decades and is still in progress from both experimental and theoretical perspectives. Accurate experimental data for level energies, spins and parities, is of utmost importance to obtain complete spectroscopic information of the calcium isotopes apart from helping us to determine the nature of interaction among the nucleons therein and set a benchmark for our theoretical calculations.

The study of $\beta$ decay is a very important tool to determine the structure of atomic nuclei. In $\beta$ decay, transitions are classified into two categories \cite{jouni}: allowed and forbidden, based on the value of orbital angular momentum ($l$) of the emitted leptons. 
Transitions with $l=0$ correspond to allowed and those with $l > 0$ to forbidden transitions.
Allowed transitions are characterised by a change in total angular momentum $\Delta J$ = 0,1 and no change in parity ($\Delta\pi$ = +1). 
Forbidden $\beta$ decays are further classified into forbidden unique (FU) and forbidden non-unique (FNU) transitions. For the $K^{th}$-FU transition, $\Delta J$ = $K+1$, while $\Delta J$ $\leq$ $K$  for the $K^{th}$-FNU transition and $\Delta\pi$ = $(-1)^K$ in both the cases. For both allowed and forbidden  transitions, $\log ft$ values, shape factors and electron spectra have been calculated with the effective value of axial-vector coupling constant g$_{\rm A}$, obtained corresponding to their mass region, earlier in $\beta$ decay studies in Refs. \cite{mika2016,mika2017,mst2006,joel12017,joel22017,VikashK}. 
Study of second-FNU $\beta^{-}$ decay of $^{24}$Na($4^{+}$) $\rightarrow$ $^{24}$Mg($2^{+}$) and $^{36}$Cl($2^{+}$) $\rightarrow$ $^{36}$Ar($0^{+}$) has been reported recently in Ref. \cite{Anil}.

Several experiments have been performed to investigate the excited states of $^{46,47}$Ca, of which, some  are  nuclear reaction based experiments  \cite{W.W.Daehnick, D.C.Williams, G.M.Crawley, M.E.Williams-Norton, P.Martin} while others are $\beta$ decay measurements \cite{B.Parsa, P.Kunz, E.K.Warburton, D.E.Alburger, T.Kuroyanagi}. Although several excited states have been identified in these experiments, experimental assignments of spin-parity to all the states could not be made.  In some instances, tentative assignments are given and in others, no assignment could be made.

Recently, an experiment has been carried out at TRIUMF in which high-statistic data from $\beta^-$ decay of ground state (g.s.) $J^{\pi} = 2^-$  of $^{46}$K into $^{46}$Ca has been measured with GRIFFIN spectrometer \cite{46Ca}, where, 42 excited states were populated and $\beta$ feeding branching ratios into excited states of $^{46}$Ca were also observed. From analysis of these data, it was possible to assign spin-parity of excited states. 
From the same experiment, $\beta^-$ decay of g.s. $J^{\pi} = 1/2^+$ of $^{47}$K into $^{47}$Ca
was studied in which spin-parity of 8 excited states were
assigned along precise measurement of half-life of $^{47}$K \cite{47Ca}.
These experiments measure simultaneously the probability of electromagnetic transitions among the states of a nucleus and $\beta$ decay feeding of these states from neighbouring nuclei. Studying them together is a real  stringent test for the nuclear shell-model.

Study of nuclear structures up to $Z = 20$ and beyond using first principles or \textit{ab initio} approaches has gained momentum over the past few years \cite{nocore_review,P.Choudhary,arch2,G.HagenPRL,T.D.Morris,Holtdrip,Anil2020}.
In Ref. \cite{G.Hagen}, a coupled-cluster (CC) method with two-nucleon (\textit{NN}) and three-nucleon (3\textit{N})  interaction from chiral effective field theory (EFT) \cite{RMP,EFT1,EFT2,EFT3} has been used to treat the many-body system of calcium isotopes. Their results \cite{G.Hagen} confirm a shell closure in $^{48}$Ca, a subshell closure in $^{52}$Ca and a weak subshell closure in $^{54}$Ca.
It has been shown that 3$N$ forces play an important role in explaining the magic number $N = 28$ in calcium isotopes \cite{J.D.Holt}.
In Ref. \cite{PRC90},  Holt \textit{el. al} have calculated energy spectra of neutron-rich calcium isotopes using many-body perturbation theory (MBPT) with \textit{NN} and \textit{NN}+3\textit{N} forces for $pf$ and $pf$g$_{9/2}$ valence spaces. They have also performed shell-model calculations using the phenomenological $pf$-shell interactions like GXPF1A \cite{M.Honma} and KB3G \cite{A.Poves} and compared  results with the experimental data. 
For $^{46}$Ca, their results are in good agreement with the experimental data, however, $4^+$ and $6^+$ states are $\sim$ 1 MeV higher for \textit{NN}+3\textit{N} interaction in $pf$g$_{9/2}$ model space than the experimental values. 
A significant spread in the spectra was obtained, indicating the need to incorporate cross-shell degrees of freedom. Further, they are unable to reproduce negative parity states and the $0^+_2$ state, which was expected to be dominated by $sd$-shell.  
Thus $sdpf$-model space is crucial for the study of these states. For $^{47}$Ca, they concluded that $pf$-valence space gives compressed spectra and the best results are obtained with $NN +3N$ forces in $pf$g$_{9/2}$ space. This establishes that the extended space $pf$g$_{9/2}$ with $3N$ forces improves the energy spectra.

In Ref. \cite{46Ca}, authors have extended the work of Holt \textit{el. al} wherein they have used non-standard valence space for $^{46}$Ca in which protons occupy $s_{1/2}d_{3/2}f_{7/2}p_{3/2}$ orbitals and neutrons occupy $s_{1/2}d_{3/2}f_{7/2}p_{3/2}p_{1/2}$ orbitals above a $^{28}$Si core. They reproduced correctly the $0^+_2$ state and also the negative parity states but obtained a significantly compressed spectra and their results showed that proton excitations from $sd$-shell are dominant. These calculations reveal that the protons are no longer inactive despite it having a robust shell-closure with the $Z = 20$.

Other \textit{ab initio} methods like in-medium similarity renormalization group (IM-SRG) targeted for a particular nucleus \cite{S.R.Stroberg} and self-consistent Green's function (SCGF) \cite{V.Soma} theory have also been applied to study the calcium isotopes.
Bhoy \textit{el. al} have used the shell-model with GXPF1Br + V$_{MU}$ interaction \cite{T.Togashi} for calculating energy spectra of $^{47-58}$Ca \cite{B.Bhoy} where they have used $pf$, $pf$g$_{9/2}$ and $pf$g$_{9/2}d_{5/2}$ model spaces to show the dependence of states on the inclusion of the higher orbitals. 

Nuclear charge radius is an important observable to describe the shell evolution in nuclei. An experiment at ISOLDE-CERN based on laser spectroscopy has been performed to measure charge radius for $^{52}$Ca \cite{Nature}, which was found to be much larger than that expected for a doubly magic nucleus. From \textit{ab initio} CC theory \cite{Nature}, these results are well reproduced. This reflects a breaking of magicity at $N = 32$ in calcium isotopes. A recent shell-model study \cite{L.Coraggio} for calcium isotopes has been reported to find the location of the drip line of calcium isotopes and showed that the calcium isotope chain is bound up to $^{70}$Ca.

In the present work we have calculated the $\log ft$ values for allowed and forbidden $\beta^-$ decay of $^{46}$K and $^{47}$K corresponding to the available experimental data of TRIUMF experiment \cite{46Ca,47Ca}, using the shell-model with $sdpf$-model space. For these calculations, we have also extracted the effective value of g$\rm_A$ using chi-square fitting method. We have also studied the energy spectra, quadrupole moment, magnetic moment and transition strengths of $^{46,47}$K and $^{46,47}$Ca isotopes. A comparison of our theoretically predicted results with the experimental ones is also presented.  Wave functions of  various states in both calcium isotopes have also been studied to determine the contribution from each configuration and also deduce the role of $sd$-model space. 

Our work is the first theoretical description of energy spectra of $^{47}$Ca corresponding to recently available experimental data \cite{47Ca}, which refined the knowledge regarding the level scheme of $^{47}$Ca. 
Our study is the first shell-model calculation of the $\log ft$ values for allowed and forbidden $\beta^{-}$ decay of $^{46,47}$K \cite{46Ca,47Ca}.

This paper is organized as follows. In Sec. II, we explain the shell-model Hamiltonian with adopted model space and theoretical formalism for allowed and forbidden nuclear $\beta$-decay. Also, we extract the effective value of g$\rm_A$ using chi-square fitting method for $\beta$ decay study. In Sec. III, we present theoretically calculated level structures and spectroscopic properties of $^{46,47}$K and $^{46,47}$Ca.
Also we have reported the $\log ft$ results of $\beta^{-}$ decay of $^{46}$K and $^{47}$K.  In Sec. IV, conclusions are drawn. 

\section{Formalism}
\subsection{Shell-model Hamiltonian}
The nuclear shell-model Hamiltonian contains a single-particle energy  and a two-nucleon interaction terms. The shell-model Hamiltonian can be written as follows
\begin{equation}
H = T + V = \sum_{\alpha}{\epsilon}_{\alpha} c^{\dagger}_{\alpha} c_{\alpha} + \frac{1}{4} \sum_{\alpha\beta \gamma \delta}v_{\alpha \beta \gamma \delta} c^{\dagger}_{\alpha} c^{\dagger}_{\beta} c_{\delta} c_{\gamma},
\end{equation}
where $\alpha = \{n,l,j,t\}$ is a single-particle state and the corresponding single particle energy is $\epsilon_{\alpha}$. $c^{\dagger}_{\alpha}$ and $c_{\alpha}$ are the creation and annihilation operators. 
$v_{\alpha \beta \gamma \delta} = \langle\alpha \beta | V | \gamma \delta\rangle $ are the antisymmetrized two-body matrix elements.

We have used the nuclear shell-model to determine the nuclear structure properties of $^{46,47}$K and $^{46,47}$Ca. It has been found that including cross-shell excitations in the valence space is necessary in order to correctly reproduce the energy spectra  of $^{46,47}$K and $^{46,47}$Ca \cite{46Ca,PRC90}. So, we have used a phenomenological SDPF-MU interaction \cite{SDPFMU} for $sdpf$-model space in the shell-model calculations with $^{16}$O as an inert core. 
In this interaction, there are seven orbitals ($d_{5/2}$, $s_{1/2}$, $d_{3/2}$, $f_{7/2}$, $p_{3/2}$, $f_{5/2}$ and $p_{1/2}$) corresponding to two major shells, available for the valence nucleons to occupy.
This allows us to calculate both the positive and negative parity states of $^{46,47}$K and $^{46,47}$Ca. 
For unnatural parity states, we have allowed either one proton or one neutron excitation from $sd$ to $pf$ shell, known as the 1p-1h excitation.

\subsection{$\beta$ decay theory for allowed and forbidden transitions} \label{beta}

A detailed theoretical formalism of nuclear $\beta$ decay has been presented in Refs. \cite{behrens1982,hfs1966}. 
A relatively simple approach in describing $\beta$ decay involves the impulse approximation, according to which, the decaying nucleon experiences just the weak interaction at the moment of decay and does not interact strongly with the remaining $A-1$ nucleons. As such, the $A-1$ nucleons are referred to as spectator nucleons with respect to the weak interaction. In the initial and final nuclear many-body states, however, the active nucleon does participate in strong interactions \cite{jouni}. The nuclear $\beta^-$ decay process is described with an effective point-like interaction vertex with an effective coupling constant $G_{\text{F}}$, known as Fermi coupling constant. The probability of the emitted electron for the $\beta^-$ decay process to lie in an energy range $W_e$\,\,to $W_e+dW_e$ is given by 


\begin{equation} \label{eq1}
\begin{split}
P(W_e)dW_e & = \frac{G_\text{F}^2}{(\hbar{c})^6}\frac{1}{2\pi^3\hbar}C(W_e)p_ecW_e(W_0-W_e)^2 \\ & \times{F_0(Z,W_e)dW_e},
\end{split}
\end{equation}
where, $W_0$ is the endpoint energy of the $\beta$ spectrum which corresponds to the maximum energy taken by the electron in a transition, the factor $F_0(Z, W_e)$ is the Fermi function, and  $Z$ is the proton number of the daughter nucleus. The $p_e$ and $W_e$ are the momentum and energy of the emitted electron, respectively and $C(W_e)$ is the shape factor that contains information about the nuclear structure.  
The partial half-life of the decay process is related to the transition probability as
\begin{eqnarray}\label{hf1}
t_{1/2}=\frac{\text{ln}(2)}{\int_{m_ec^2}^{W_0}{P(W_e)dW_e}},
\end{eqnarray}
where $m_e$ is the rest mass of the electron and the integral in the denominator is called the decay rate or transition probability per unit time. To simplify this integration, we introduce the dimensionless quantities $w_0=W_0/m_ec^2$, $w_e=W_e/m_ec^2$, and $p=p_ec/m_ec^2=\sqrt{(w_e^2-1)}$, which are kinematical quantities divided by the electron rest mass.  
 Now, the partial half-life is related to shape function as 
 \begin{equation}
 	ft_{1/2}={\kappa}  ,
 \end{equation}
where, $f$ is the dimensionless integrated shape function and  the value  of the constant $\kappa$ \cite{Patrignani} is

\begin{eqnarray}
\kappa=\frac{2\pi^3\hbar^7\text{ln(2)}}{m_e^5c^4(G_\text{F}\text{cos}\theta_\text{C})^2}=6289~\mathrm{s},
\end{eqnarray}
where, $\theta_\text{C}$ is the Cabibbo angle.

The dimensionless integrated shape function $f$  can be written as 
 
\begin{eqnarray} \label{tc}
f=\int_1^{w_0}C(w_e)pw_e(w_0-w_e)^2F_0(Z,w_e)dw_e.
\end{eqnarray}
 
For allowed (Gamow-Teller) transition, shape factor $C(w_e)$ can be expressed as
\begin{equation}
C(w_e)  =B(GT) =\frac{g_{\rm A}^{2}}{2J_{i}+1}|\mathcal{M}_{\rm GT}|^2,
\end{equation}

where, $\mathcal{M}_{\rm GT}$ is the Gamow-Teller nuclear matrix element (NME) of \bm{$\sigma\tau$} and $B(GT)$ is reduced transition probability (details of which can be found in Ref. \cite{jouni}), g$_{\rm A}$ is the axial-vector coupling constant and $J_{i}$ is the  angular momentum of initial state. 
For such a transition, the reduced half-life  is calculated by multiplying the half-life with the phase-space factor (Fermi integral) given by
\begin{equation}\label{f0}
	f_{0}= \int_1^{w_0}pw_e(w_0-w_e)^2F_0(Z,w_e)dw_e.
\end{equation}
We then take the logarithm of the \textit{ft value} also known as the reduced half-life or comparative half-life, as   the \textit{ft value}  is  generally a large quantity.

\begin{equation*}
	\mbox{log} ft \equiv \mbox{log}(f_{0}t_{1/2}~s).
\end{equation*}

The shape factor $C(w_e)$ for forbidden transition has the form of 

\begin{align} 
	\begin{split}
		C(w_e) & = \sum_{k_e,k_\nu,K}\lambda_{k_e} \Big[M_K(k_e,k_\nu)^2+m_K(k_e,k_\nu)^2 \\
		&-\frac{2\gamma_{k_e}}{k_ew_e}M_K(k_e,k_\nu)m_K(k_e,k_\nu)\Big],
		\label{eq2}
	\end{split}
\end{align}
where, $k_e$ and $k_\nu$ (= 1, 2, 3,...) are related to the partial-wave expansion of the leptonic wave functions, and $K$ is the order of forbiddeness of the transition. The entities $M_K(k_e,k_\nu)$ and $m_K(k_e,k_\nu)$ are written in terms of NME and leptonic phase-space factors \cite{behrens1982,mika2017}. 
The quantities
\begin{align*}
	\gamma_{k_e} & =\sqrt{k_e^2-(\alpha{Z})^2}&\mbox{and} \,\,\,\, \lambda_{k_e} & ={F_{k_e-1}(Z,w_e)}/{F_0(Z,w_e)},
\end{align*}

where, $\lambda_{k_e}$ is the  Coulomb function, in which ${F_{k_e-1}(Z,w_e)}$ is the generalized Fermi function \cite{mst2006,mika2017}. The $\log ft$ values for forbidden transitions are calculated in a similar manner as that in the case of allowed transitions by using the phase-space factor corresponding to the concerned transitions (more details can be found in Ref. \cite{behrens1982}). The NME contain all the information about nuclear-structure and can be expressed as

\begin{align}
\begin{split}
^{\rm V/A}\mathcal{M}_{KLS}^{(N)}(pn)(k_e,m,n,\rho)& \\ =\frac{\sqrt{4\pi}}{\widehat{J}_i}
\sum_{pn} \, ^{\rm V/A}m_{KLS}^{(N)}(pn)(&k_e,m,n,\rho)(\Psi_f|| [c_p^{\dagger}
\tilde{c}_n]_K || \Psi_i),
\label{eq:ME}
\end{split}
\end{align}
where, the summation, in Eq. \ref{eq:ME}, runs over the proton $(p)$ and neutron $(n)$ single-particle states with ${\widehat{J}_i}=\sqrt{2J_i+1}$. The quantity m is the total power of ($m_eR/\hbar$), ($W_eR/\hbar$) and $\alpha$Z; n is the total power of ($W_eR/\hbar$) and $\alpha$Z; $\rho$ is the power of $\alpha$Z.
The functions ${^{V/A}m_{KLS}^{(N)}}(pn)(k_e,m,n,\rho)$ are the single-particle matrix elements (SPMEs) that does not depend on the nuclear models and $(\Psi_f|| [c_p^{\dagger}\tilde{c}_n]_K || \Psi_i)$ are the one-body transition densities (OBTDs) between the initial $(\Psi_i)$ and final $(\Psi_f)$ states.
SPMEs expressions, used in this work, are taken from \cite{mika2017,mst2006} and OBTDs are computed from the shell-model code NUSHELLX \cite{nushellx}.

\subsection{Quenching factor}
\vspace{-0.7cm}
\begin{table}[!ht]
	\caption{Comparison of computed $\log ft$ values and average shape factors for Gamow-Teller and first-forbidden transitions with experimental data \cite{46Ca,47Ca}. Only confirmed transitions for $^{46}$K $\rightarrow$ $^{46}$Ca and $^{47}$K $\rightarrow$ $^{47}$Ca are presented. Calculated values are obtained by using quenching factor q = 1.00 and enhancement factor $\epsilon_{\text{MEC}}$ = 1.6 for $0^-$ transitions.}   
	\label{Cbar}
	\begin{ruledtabular}
		\begin{tabular}{lccccc}
			\multicolumn{2}{c}{~~~~~Transition}	& \multicolumn{2}{c}{~~~~~$\log ft$} &\multicolumn{2}{c}{~~~~~${\overline{(C(w_{e}))}}^{1/2}$} \\
			\cline{1-2}
			\cline{3-4}
			\cline{5-6}
			Initial& Final & Expt. &Theory   & Expt. &Theory  \T\B\\\hline\T\B
			
			$^{46}$K$(2^-)$ & $^{46}$Ca$(2^+_1)$	& 6.92	& 6.439		&10.62  &18.47\\ \T\B
			& $^{46}$Ca$(0^+_2)$	& 10.27	& 9.525		&0.22	&0.53\\ \T\B
			& $^{46}$Ca$(4^+_1)$	& 9.70	& 7.405		&0.43	&6.075\\ \T\B
			& $^{46}$Ca$(2^+_2)$	& 8.01	& 8.298		&3.03	&2.17\\ \T\B
			& $^{46}$Ca$(3^-_1)$	& 8.04	& 5.755		&0.0076	&0.105\\ \T\B
			& $^{46}$Ca$(2^+_3)$	& 8.13	& 6.879		&2.64	&11.135\\ \T\B
			& $^{46}$Ca$(4^+_2)$	& $>$11.5	& 8.868		&0.054	&1.13\\ \T\B						
			$^{47}$K$(1/2^+)$ & $^{47}$Ca$(3/2^-_1)$	&$>$6.5	& 7.831		&17.22	&3.72\\ \T\B
			& $^{47}$Ca$(3/2^+_1)$	& 5.457	& 5.690		&0.1482	&0.11\\ \T\B
			& $^{47}$Ca$(1/2^+_1)$	& 4.807	& 4.386		&0.3132	&0.51\T\B  
		\end{tabular}
	\end{ruledtabular}	
\end{table}
The bare values of the two coupling constants, g$\rm _V$ = 1.0 (vector coupling constant) and g$\rm _A$ = 1.27 (axial-vector coupling constant), associated with the $\beta$ decay theory are determined from conserved vector current (CVC) and partial conserved axial-vector current (PCAC), respectively.
In the nuclear shell-model calculations, 
 the renormalizations due to the truncation of model space and possible meson-exchange current contributions are taken into account by using effective values of the coupling constants.
 In this paper, we have calculated the quenching factor for g$\rm _A$. The averaged shape factor \cite{behrens1982} is defined as 

\begin{equation}
	\overline{C(w_{e})} = f/f_{0}\,.
\end{equation}
The expressions of $f$ and $f_{0}$ are written in Eq. \ref{tc} and \ref{f0}. For Gamow-Teller transitions, $\overline{C(w_{e})}$ is $B(GT)$ and that for first-forbidden transitions is \cite{enhancement,PRC85}
\begin{equation}
	\overline{C(w_{e})} (fm^2) =\frac{6289 \lambdabar_{\text{Ce}}^2}{ft} = \frac{9378 \times 10^5}{ft},
\end{equation} 

where, $\lambdabar_{\text{Ce}}$ is the reduced Compton wavelength of electron.
For first-forbidden $\beta$ decays, there are six operators for NME, given by
\begin{align}
	\begin{split}
	O(0^-) &: \text{g}_{\rm A}(\bm{\sigma} \cdot \bm{p_e}),\,\,\, \text{g}_{\rm A}(\bm{\sigma} \cdot\bm{ r}) \\
	O(1^-) &: \text{g}_{\rm V}{\bm{p_e}},\,\,\, \text{g}_{\rm A}(\bm{\sigma}\, \times \,\bm{ r}), \,\, \text{g}_{\rm V}\bm{r} \\
	O(2^-) &: \text{g}_{\rm A}[\bm{\sigma r}]_2\,,
	\end{split}
\end{align}
where, {\bm {$r$}} and {\bm {$p_e$} } are the radial coordinate and electron momentum, respectively. The operators $O(0^-)$, $O(1^-)$ and $O(2^-)$ correspond to $\Delta J$ = 0, 1 and 2 transitions, respectively.
 The operator \bm{$\sigma$} $\cdot$ \bm{$p_e$} is the axial-charge nuclear matrix element $\gamma_5$. For first-forbidden non-unique $J^+ \leftrightarrow J^-$ transitions, $\gamma_5$  NME is enhanced due to nuclear medium effects in the form of meson-exchange currents. 
 This enhancement factor in $\gamma_5$ is represented by $\epsilon_{\text{MEC}}$, defined as the ratio of the axial-charge matrix element to its impulse-approximation value.
  The one-body impulse NME $M_0^{T}$ =  $<$ J$_f$ $||$ -g$_A$ $\frac{1}{M} \vec{\sigma}\cdot\vec{p}_e$ $t_{-}$ $||$ J$_i$ $>$, where M is the nucleon mass and $t_{-}|n> =|p>$, is multiplied by an effective coupling constant, $\epsilon_{\text{MEC}}$, to take account of the contributions from the meson-exchange currents \cite{Towner,WTB}. 
  We have taken the value of $\epsilon_{\text{MEC}}$ as 1.6 which is consistent with Ref. \cite{enhancement,meson}.

\begin{figure}
	\includegraphics[width=\columnwidth]{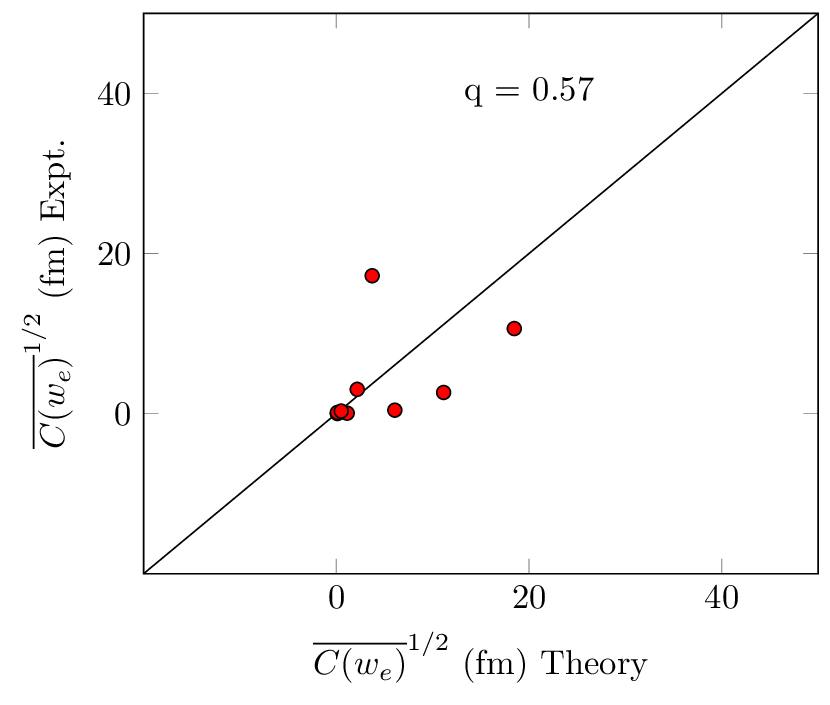}
	\caption{Comparison of calculated and experimental average shape factors for Gamow-Teller and first-forbidden transitions to obtain the quenching factor. For extraction of the quenching factor, we have used only confirmed transitions of $^{46,47}$Ca.}
	\label{Average_shape_factor}
\end{figure}

In Table \ref{Cbar}, we have shown theoretical and experimental $\log ft$ values and averaged shape factors corresponding to transitions $^{46}$K $\rightarrow$ $^{46}$Ca and $^{47}$K $\rightarrow$ $^{47}$Ca. These calculations are done by using free coupling constants (quenching factor q = 1.0). 
We have also included mesonic enhancement factor $\epsilon_{\text{MEC}}$ = 1.6 for the matrix element $M_0^{T}$ in 0$^{-}$ transitions.
 In the present calculations, we have also included next-to-leading-order terms of $\beta$ decay shape function, similar to \cite{mika2017,mika2016}.  
In Fig. \ref{Average_shape_factor}, theoretical and experimental ${\overline{C(w_{e})}}^{1/2}$ values are plotted. We have extracted the effective value of g$\rm _A$ = q $\times$ g$\rm _A^{\,\text{free}}$ = 0.72 corresponding to quenching factor q = 0.57 from the chi-square fitting method.


\section{Results and Discussion}

\begin{figure*}
	\subfloat[\label{A1}]{\includegraphics[width=0.48\textwidth]{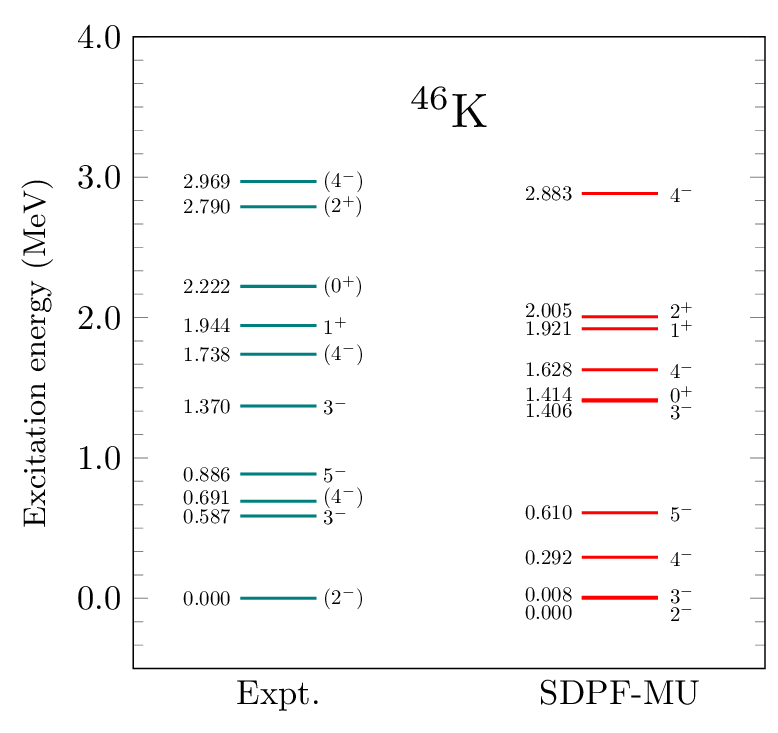}}\hfill
	\subfloat[\label{B1}]{\includegraphics[width=0.48\textwidth]{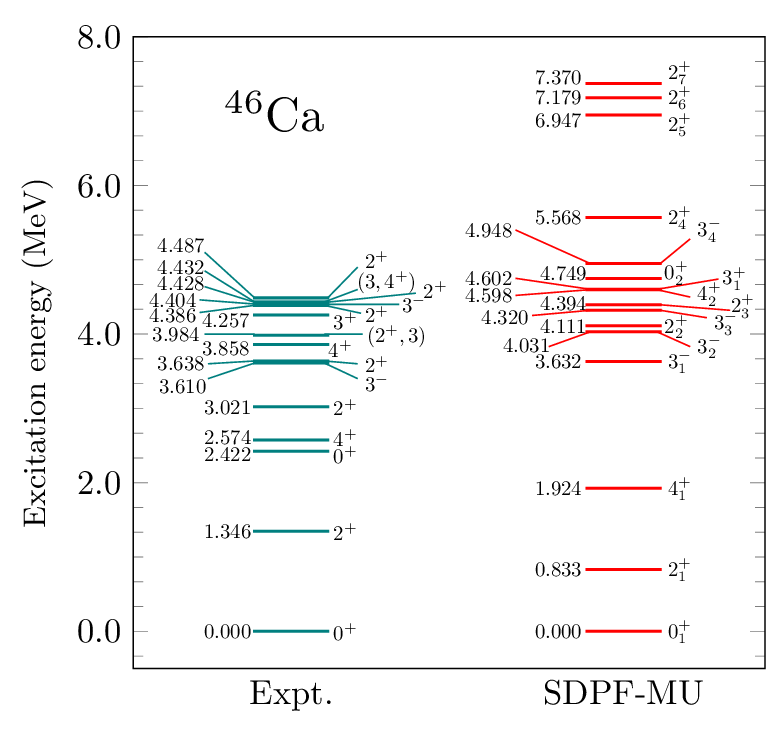}}
	\caption{Comparison of theoretical and experimental energy spectra of $^{46}$K and $^{46}$Ca. Experimental data are taken from Refs. \cite{NNDC,47Ca}.}
	\label{46K46Ca}	
\end{figure*}

\begin{figure*}
	\subfloat[\label{A2}]{\includegraphics[width=0.48\textwidth]{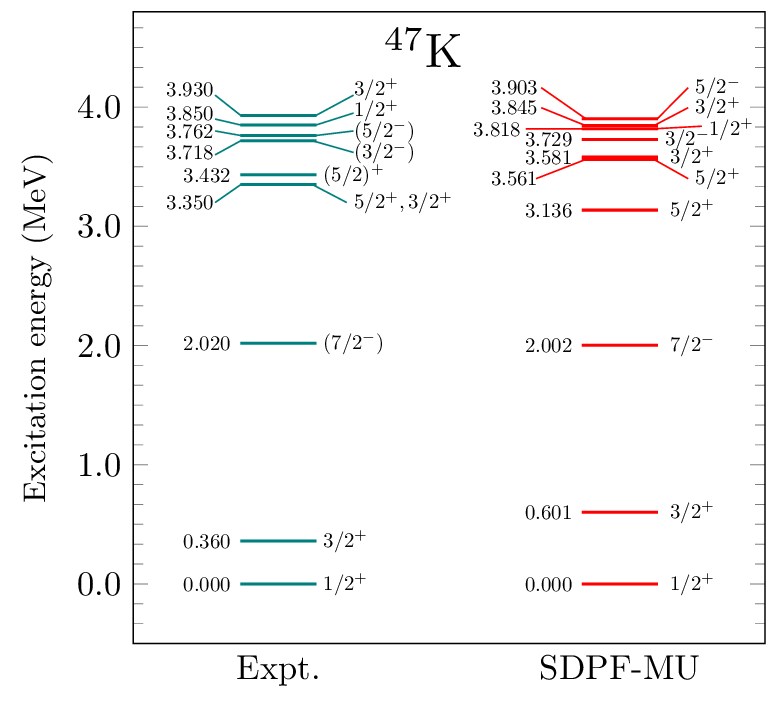}}\hfill
	\subfloat[\label{B2}]{\includegraphics[width=0.48\textwidth]{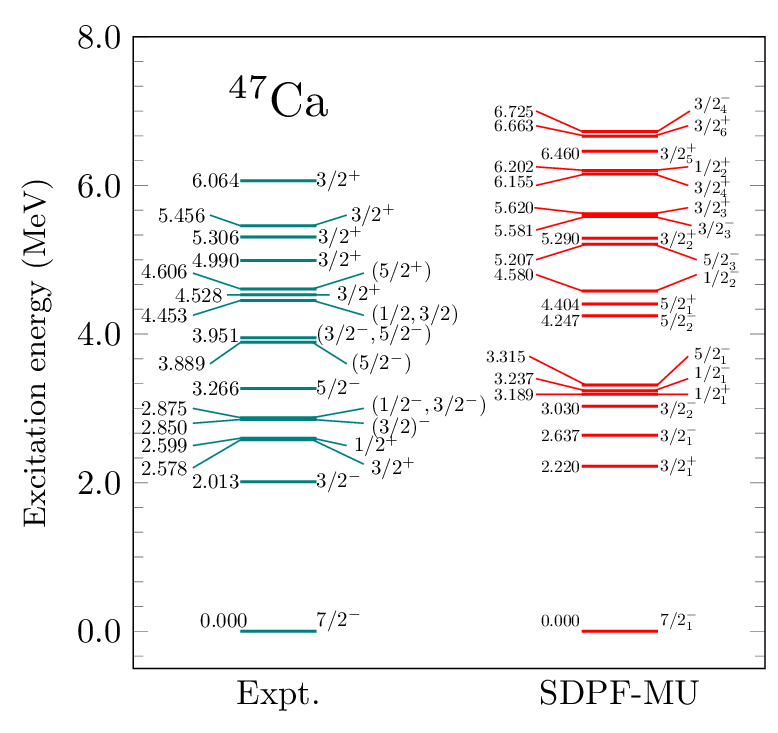}}
	\caption{Comparison of theoretical and experimental energy spectra of $^{46}$K and $^{46}$Ca. Experimental data are taken from Refs. \cite{NNDC,47Ca}.}
	\label{47K47Ca}	
\end{figure*}

We have performed shell-model calculations to study nuclear structures of $^{46,47}$K and $^{46,47}$Ca in $sdpf$-model space. 
Low-lying energy spectra of $^{46}$K and $^{46}$Ca obtained from the shell-model are compared with the experimental data in Fig. \ref{46K46Ca}. 
The experimental spin-parity of g.s. is tentative for $^{46}$K. Our theoretical calculations predict the g.s. as $2^{-}$. The g.s. of $^{46}$K has the configuration of $\pi(d_{5/2}^6s_{1/2}^2d_{3/2}^3)$ $\otimes$ $\nu(d_{5/2}^6s_{1/2}^2d_{3/2}^4f_{7/2}^7)$ with 70.24\% probability.
In Fig. \ref{47K47Ca}, low-lying energy spectra corresponding to $^{47}$K and $^{47}$Ca are shown and also compared with the experimental data. The shell-model calculations with SDPF-MU interaction correctly reproduce the g.s. spin-parity for $^{47}$K. The g.s. configuration of $^{47}$K is $\pi(d_{5/2}^6s_{1/2}^1d_{3/2}^4)$ $\otimes$ $\nu(d_{5/2}^6s_{1/2}^2d_{3/2}^4f_{7/2}^8)$ with probability of 71.21\%. The calculated energy spectra of $^{47}$K is in good agreement with the experimental one.

\begin{table}[!ht]
\caption{Comparison of theoretical and experimental reduced electric quadrupole transition probabilities from excited states to g.s. in $^{46,47}$K  and $^{46,47}$Ca. Effective charges are taken as $e_p = 1.35e$ and $e_n = 0.35e$.  Experimental values are taken from Refs. \cite{NNDC,Qandmag}.}
\label{B(M1)_B(E2)}
\begin{ruledtabular}
		\begin{tabular}{lccccc}
		& & \multicolumn{2}{c}{$B(E2) (W.u.)$}  \T\B\\
		\cline{3-4}
		Nuclei & Transition &  Expt. & SDPF-MU \T\B\\\hline
		$^{46}$K & $4^-_1 \rightarrow 2^-_1$ & NA & 2.70 \T\B\\
		$^{46}$Ca & $2^+_1 \rightarrow 0^+_1$ & 3.63 & 0.40 \T\B\\
        $^{47}$K & $5/2^+_1 \rightarrow 1/2^+_1$& NA  & 2.5\T\B\\
		$^{47}$Ca & $3/2^-_1 \rightarrow 7/2^-_1$& $< 0.28$  & 0.2\T\B\\
		& $3/2^-_2 \rightarrow 7/2^-_1$& NA & 0.30\T\B\\
		\end{tabular}
	\end{ruledtabular}
\end{table}


We have computed electromagnetic transition strengths $B(E2)$ for transitions from  excited states to g.s. in $^{46,47}$K and $^{46,47}$Ca, which are shown in Table \ref{B(M1)_B(E2)}. In case of $^{46}$Ca, $B(E2)$ value for $2^+_1 \rightarrow 0^+_1$ transition is 0.4 W.u. while experimental value is 3.63 W.u.,  which is quite far from theory but $B(E2$ $3/2^-_1 \rightarrow 7/2^-_1$) in $^{47}$Ca is 0.2 W.u.  which agrees with the experimental value ($< 0.28$ W.u.). 
We have also predicted experimentally unknown transition strength $B(E2; 3/2^-_2 \rightarrow 7/2^-_1)$ for $^{47}$Ca.

In Table \ref{MU_Q}, we have reported quadrupole and magnetic moments obtained from the shell-model with SDPF-MU interaction for selected states of $^{46,47}$K and $^{46,47}$Ca. We have then compared these values with the experimental ones \cite{NNDC,Qandmag} and found that the magnetic moments of $2^+$, $1/2^+$ and $7/2^-$ states are consistent with experimental values for $^{46}$Ca, $^{47}$K and $^{47}$Ca, respectively, for $g_s^{eff}$ = 0.85 $g_s^{free}$. Also, we obtain magnetic moment of g.s. $2^-$ for $^{46}$K 
as -0.478 (-0.585) for $g_s^{eff}$ = 0.85 (0.7)$g_s^{free}$.
As the quenching factor for $g_s$ is found to be 0.9 in $fp$-shell for the GXPF1 interaction \cite{Honma} and 0.92-0.93 in $sd$-shell for the USDB interaction \cite{Richter}, the value of 0.85 is a reasonable one for SDPF-MU.
 We have  predicted quadrupole and magnetic moments for several states of $^{46,47}$Ca from the shell-model which are experimentally not yet measured.

So, our energy spectra and  magnetic moments
 corresponding to the adopted interaction are in a reasonable agreement with the experimental data. Thus, SDPF-MU interaction is a suitable interaction for calculating nuclear properties of these nuclei. It means that we can use the wave functions, obtained from SDPF-MU interaction, to compute the NME needed for the  $\log ft$ calculations.

\begin{table}[!ht]
	
	\caption{Comparison of  theoretical and experimental quadrupole and magnetic moments of selected states for $^{46,47}$K and $^{46,47}$Ca. The
effective charges are taken as $e_p = 1.35e$ and $e_n = 0.35e$. 
 For the magnetic moment calculations two sets of $g_s^{eff}$ are employed: $g_s^{eff}$ = 0.7 $g_s^{free}$  and $g_s^{eff}$ = 0.85 $g_s^{free}$, separated by a solidus (/). 
 Experimental data are taken from Refs. \cite{NNDC,Qandmag}.}
	\label{MU_Q}
	\begin{ruledtabular}
		\begin{tabular}{lccccc}
			&   &\multicolumn{2}{c}{$\mu(\mu_N)$} &\multicolumn{2}{c}{$Q(eb)$}\\
			\cline{3-4}
			\cline{5-6}
			Nuclei & $J^\pi_f$  &Expt.& SDPF-MU    & Expt. & SDPF-MU\T\B \\\hline
			$^{46}$K & $2^-_{g.s.}$ & -1.051(6) & -0.585 / -0.478 & NA & 0.061 \T\B\\
			$^{46}$Ca & $2_1^+$ & -0.52(12) & -0.425 / -0.517 & NA & -0.043 \T\B\\
			& $3_1^+$ & NA & -1.009 / -1.225 & NA & 0.002\T\B\\
			&$4_1^+$ & NA & -1.176 / -1.429& NA & -0.015 \T\B\\
			$^{47}$K & $1/2^+_{g.s.}$ & 1.933(9) & 1.583 / 1.953 & - & -  \T\B\\
			$^{47}$Ca & $7/2_1^-$ &  -1.38(3) & -1.026 / -1.246 & 0.021(4) & 0.051\T\B\\
			& $5/2_1^-$ &  NA & -0.734  /  -0.892 & NA & -0.014\T\B\\
			& $3/2_1^-$ &  NA & -1.060  / -1.287 & NA & -0.026\T\B\\
		\end{tabular}
	\end{ruledtabular}
\end{table}

\begin{table*}[htp]
	\caption{ Computed $\log ft$ values for the allowed, FU and FNU $\beta^-$ decay from $^{46}$K (2$^-$) $\rightarrow$ $^{46}$Ca (J$_f$).  The results are obtained by using the quenching factor of q=0.57 for g$_{A}$. Experimental and theoretically predicted spin-parities are also mentioned. 
		Experimental excitation energies ($E$), $\beta^{-}$ branching ratios and $\log ft$ values are taken from Ref. \cite{46Ca}.}
	\label{46Ca_logft}
\begin{ruledtabular}
	\begin{tabular}{lcccccc}
& & &  &&\multicolumn{2}{c}{~~~~~$\log ft$}  \\
\cline{6-7}
$J^\pi_f$ (Expt.) &  $J^\pi_f$ (SM)& $E$ (keV) & Mode of decay  &$~~~\beta^-$ branching   & Expt. & Theory   \T\B\\\hline
2$^+$ &		2$^+_1$  &1346.00(14) & 1st FNU &27.4(21) & 6.92(4) &6.862 	\T\B\\
0$^+$ &		0$^+_2$  &2421.98(22) & 1st FU  &0.38(1)  & 10.27(2)&10.017	\T\B\\
4$^+$ &		4$^+_1$  &2574.53(16) & 1st FU  &1.15(14) & 9.70(6) &7.898	\T\B\\
2$^+$ &		2$^+_2$  &3020.67(14) & 1st FNU & 0.55(20)& 8.01(16)&8.772	\T\B\\
3$^-$ &		3$^-_1$  &3610.51(15) & Allowed &0.28(14) & 8.04(22)& 6.248 \T\B\\
2$^+$ &		2$^+_3$  &3638.38(15) & 1st FNU &0.22(2)  & 8.13(4)& 7.370  \T\B\\
4$^+$ &		4$^+_2$  &3857.51(17) & 1st FU &$<$0.003  &  $>$11.5&9.361	\T\B\\
(2$^+$,3) & 2$^+_4 \left.\middle/ \, \text{3}^-_2\right.$ &3984.07(16) & 1st FNU(2$^+_4) \left.\middle/ \, \text{Allowed}(3^-_2) \right.$& 0.47(3)   &7.63(3)  & $\left. 9.080 \middle/  6.009 \right.$  \T\B\\
3$^+$ &		3$^+_1$  &4257.26(15) & 1st FNU & 0.092(9) & 8.20(5)&  10.512 \T\B\\
2$^+$ &		2$^+_5 \left.\middle/ \,\text{2}^+_4\right.$ &4385.91(15) & 1st FNU &4.39(16)  & 6.44(2) 	&$\left. 9.132 \middle/ 9.137\right.$	\T\B\\
3$^-$ &		3$^-_2 \left.\middle/ \, \text{3}^-_3\right.$ &4404.48(15) & Allowed & 5.09(22) & 6.37(2) & $\left. 6.009 \middle/ 6.040\right.$	\T\B\\
2$^+$ &		2$^+_6 \left.\middle/ \, \text{2}^+_5\right.$ & 4428.02(14) & 1st FNU & 0.017(14)& 8.80(40)& $\left. 9.941 \middle/9.133\right.$	\T\B\\
(3,4$^+$) & 3$^-_3 \left.\middle/ \, \text{3}^-_4\right.$ &4432.05(16) & Allowed & 2.59(7)   & 6.65(2) & $\left. 6.040 \middle/ 7.779\right.$\T\B\\
2$^+$ &   2$^+_7 \left.\middle/ \, \text{2}^+_6\right.$  & 4487.05(16) & 1st FNU & 0.46(2)  & 7.36(2) & $\left. 9.173 \middle/9.947\right.$	\T\B\\	 

	\end{tabular}
\end{ruledtabular}
\end{table*}

\begin{table*}[!ht]

\caption{Computed $\log ft$ values for the allowed, FU and FNU  $\beta^-$ decay from $^{47}$K (1/2$^+$) $\rightarrow$ $^{47}$Ca (J$_f$).  The results are obtained by using the quenching factor of q=0.57 for g$_{A}$. Experimental and theoretically predicted spin-parities are also mentioned. Experimental excitation energies ($E$), $\beta^{-}$ branching ratios and $\log ft$ values are taken from Ref. \cite{47Ca}. }
\label{47Ca_logft}
\begin{ruledtabular}
\begin{tabular}{lcccccc}
& & &&&\multicolumn{2}{c}{~~~~~$\log ft$}   \\
\cline{6-7}
$J^\pi_f$ (Expt.)&  $J^\pi_f$ (SM)& $E$ (keV)  & Mode of Decay&$\beta^-$ branching & Expt. & Theory  \T\B\\\hline
 $3/2^-$ &    $3/2^-_1$     & 2013.4(3)  &1st FNU  &$<2.9$       &$>$6.5   &7.956	\T\B\\
 $3/2^+$ &    $3/2^+_1$     & 2578.1(3)  &Allowed  &18.4(3)      &5.457(8) &6.183	\T\B\\
 $1/2^+$ &    $1/2^+_1$     & 2599.2(3)  &Allowed  &80(2)        &4.807(9) &4.879	\T\B\\
$(3/2)^-$ &   $3/2^-_2$     & 2849.9(3)  &1st FNU  &0.013(1)     &8.46(4)  &8.201 \T\B\\
 $(1/2^-,3/2^-)$ & 1/2$^-_1$& 2874.7(3)  &1st FNU  &0.015(3)     &8.40(9)  &8.487 \T\B\\
 $5/2^-$ &      $5/2^-_1$   & 3266.3(3)  & 1st FU  &0.0038(2)    &10.28(2) &8.485	\T\B\\
 $(5/2^-)$ &     $5/2^-_2$  & 3889.0(3)  &1st FU   &$<0.0017$    &$>$10.1  &10.045 \T\B\\
$(3/2^-,5/2^-)$ & 3/2$^-_3$ & 3951.2(3)  & 1st FNU &0.010(5)     &7.9(2) or 9.3(2) & 9.896\T\B\\
$(1/2,3/2)$&  1/2$^-_2$     & 4453.1(4)  & 1st FNU  &0.036(5)    &6.99(6) &8.159 \T\B\\
		   &  1/2$^+_2$     & 4453.1(4)  & Allowed  &0.036(5)    &6.99(6) &5.811 \T\B\\
$3/2^+$  &   $3/2^+_2$      & 4528.3(3)  &Allowed &0.38(1)       &5.91(1) &5.454\T\B\\
$(5/2^+)$ &  $5/2^+_1$      & 4606.1(3)  &2nd FNU &$<0.0027$     &$>$8.0  &11.751\T\B\\
$3/2^+$  &   $3/2^+_3$      & 4989.5(3)  &Allowed &0.246(6)      &5.64(1) &5.233	\T\B\\
$3/2^+$  &   $3/2^+_4$      & 5305.9(3)  &Allowed &0.291(7)      &5.20(1) &5.534 \T\B\\
$3/2^+$  &   $3/2^+_5$      & 5456.1(3)  &Allowed &0.38(1)       &4.87(1) &5.668\T\B\\
$3/2^+$  &   $3/2^+_6$      & 6064.2(4)  & Allowed&0.0049(2)     &5.56(2) &5.382 \T\B\\

\end{tabular}
\end{ruledtabular}
\end{table*}

In Table \ref{46Ca_logft}, we have presented the calculated $\log ft$ values for $\beta^-$ decay from 2$^-$ of $^{46}$K to various excited states of $^{46}$Ca. There are two possible spin-parities: 2$^+_4$ and 3$^-_2$, corresponding to the state at 3984.07(16) keV. Accordingly, the spin-parity and $\log ft$ values of the subsequent states are determined. For instance, if the state at 3984.07(16) keV is 2$^+_4$, then the spin-parity and $\log ft$ value of the state at 4385.91(15) keV are 2$^+_5$ and 9.132, respectively. On the other hand, if the state at 3984.07(16) keV is considered to be 3$^-_2$, then, the 4385.91(15) keV state would be 2$^+_4$ with a $\log ft$ value of 9.137.
We have also listed the experimentally obtained  spin-parity $J_f^{\pi}$, excitation energy, branching ratio and $\log ft$ values of final states. Further, we have also presented the corresponding modes of decay. Transition from initial state to final state could be either allowed or FU or FNU, based on the selection rules of $\beta$ decay. 

In computing the $\log ft$, we have used the effective value of g$_{\rm A}$ = 0.72. 
 Experimental values are used for the $\beta$ decay (ground-state-to-ground-state) Q values for $^{46}$K and $^{47}$K as well as for the excitation energies in $^{46}$Ca and $^{47}$Ca for the evaluations of the log $ft$ values.  The Q values of 7.716 and 6.632 MeV are used for $^{46}$K and $^{47}$K, respectively.

The g.s. spin-parity of $^{46}$Ca is well reproduced with the SDPF-MU interaction. The energy difference between experimental and theoretical $2_1^+$ is 0.513 MeV,  as can be seen in Fig. \ref{B1} and the calculated $\log ft$ value is 6.862 which is very close to the experimental value of 6.92(4). 
Transition from $2^-$ of $^{46}$K to $0^+_2$ of $^{46}$Ca is of the 1st FU type for which calculated $\log ft$ value is 10.017, which agrees well with the experimental value of 10.27(2). The difference between the theoretically and experimentally obtained energies of $3^-_1$ is 0.022 MeV which is very small. 
Spin-parity corresponding to the state with energy 3.984 MeV is unconfirmed. For this state, experimentally, it is found that the spin-parity  based on  the $\log ft$ value is constrained to be either  $2^+$ or $3^+$ or $3^-$. If we assume this state to be $2^+$, then the $\log ft$ value corresponding to the $\beta^-$ decay to this state is 9.080. By assuming this state to be $3^+$, we obtain $\log ft$ = 10.457.  If allowed transition is assumed then this state would be $3^-$ along with $\log ft$ value of 6.009. The energy difference between calculated and experimental value for this state is 47 keV, as shown in Fig. \ref{B1}. Thus, shell-model calculation with the SDPF-MU interaction establish this state to be either $3^-$ or $2^+$.
Spin-parity of the state at 4.432 MeV is likewise unconfirmed from experiments. So, we have taken all possibilities into account and calculated corresponding $\log ft$ values. Theoretically, we predict this state to be $3^-$ on the basis of $\log ft$ value.

In Table \ref{47Ca_logft}, we have reported calculated $\log ft$ results for the allowed, FU and  FNU $\beta^-$ decay from 1/2$^+$ of $^{47}$K to various excited states of $^{47}$Ca.
The $7/2^-$ state is the g.s. of $^{47}$Ca which is correctly reproduced with SDPF-MU interaction. 
The order of experimental $3/2^-_1$ and $3/2^+_1$ states  is reversed in theoretical prediction by shell-model,  as shown in Fig. \ref{B2}. 
The calculated $\log ft$ for the $3/2^-_{2}$ state is 8.201 which is very near to experimental $\log ft$ of  8.46(4) for $(3/2)^-$ state.
The experimental $(3/2)^-$ state is at 2.850 MeV while theoretical value of energy for this state is 3.030 MeV which means shell-model confirm this state to be $3/2^-$.
Next unconfirmed state is at 2.875 MeV, which is constrained to be $(1/2^-,3/2^-)$, experimentally. If this state is considered to be $3/2^-$ then calculated $\log ft$ is 9.846, while for $1/2^-$, $\log ft$ is 8.487 which is very close to experimental $\log ft$  \textit{i.e.} 8.40(9). Also excitation energy for $1/2^-$ is 3.237 MeV which is quite close to the experimental value. So, we predict a state at 2.875 MeV to be $1/2^-$.
Experimental first $5/2^-$ with $\log ft$ value of 10.28(2) 
 can be assigned as shell-model $5/2_1^-$ state. We also get a confirmation between theory and experiment for $(5/2)^-$ state at 3.889 MeV. For this state calculated $\log ft$ is 10.045, while experimental value is $>$ 10.1.

For a state at 3.951 MeV in $^{47}$Ca, there are two possibilities, see in Fig. \ref{B2}. It could be either $3/2^-$ or $5/2^-$. If this state is assumed as $5/2^-$ then there occurs 1st FU transition and calculated $\log ft$ is  13.170 which is far from experimental $\log ft$ value of 9.3(2). If we consider this state as $3/2^-$ then transition is to be 1st FNU type and calculated $\log ft$ value is 9.896,  while the experimental value is 7.9(2).  The state is more likely to be $3/2^-$.
There are four possibilities of spin-parity for a state at 4.453 MeV excitation energy. This could be either $1/2^-$, $1/2^+$, $3/2^-$ or $3/2^+$. The calculated $\log ft$ values are 8.159, 5.811, 10.815 and 5.454 for $1/2^-$, $1/2^+$, $3/2^-$ and $3/2^+$ states, respectively. Also calculated excitation energies with SDPF-MU interaction for these states are 4.58, 6.202, 6.725 and 5.290 MeV, respectively. So,  the $\log ft$ values favour this state as either $1/2^-$ or $1/2^+$.
The experimental $(5/2)^+$ state is at 4.606 MeV, while the calculated value of  $\log ft$ for this state is 11.751 (experimental one is $>$ 8.0) and calculated energy is 4.404 MeV. Thus, shell-model results support this states as $5/2^+$.

\begin{figure}[htb!]
	\includegraphics[width=8.7cm,height=8cm]{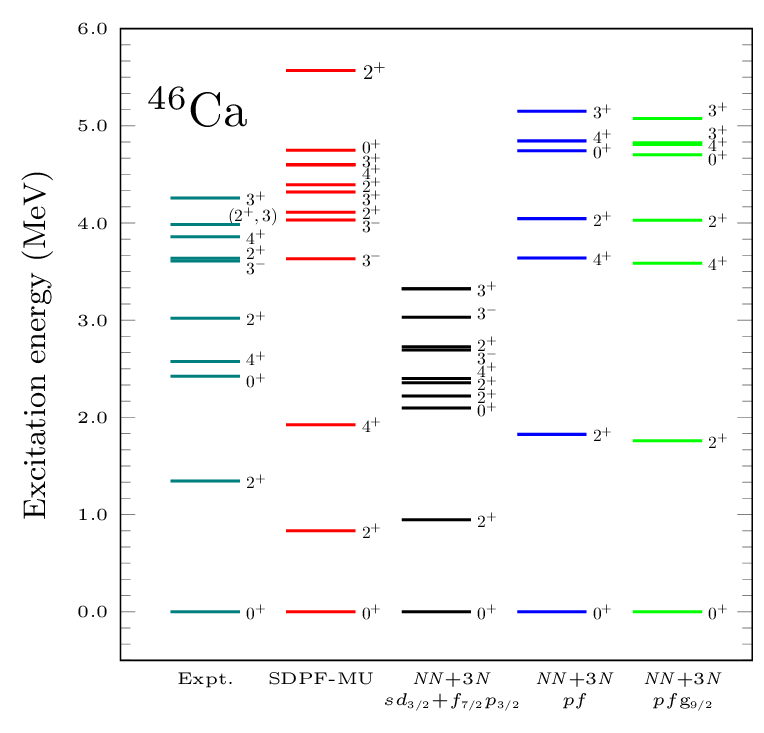}
	\caption{Comparison of excitation energies of low-lying excited states in $^{46}$Ca obtained from SDPF-MU interaction with previously obtained results in $sd_{3/2}+f_{7/2}p_{3/2}$, $pf$ and $pf$g$_{9/2}$ valence spaces using \textit{NN}+3\textit{N} forces \cite{46Ca},\cite{PRC90}. Experimental data are also shown in first column for comparison.}
	\label{Comparison1}
\end{figure}

\begin{figure}
	\includegraphics[width=8.7cm,height=8cm]{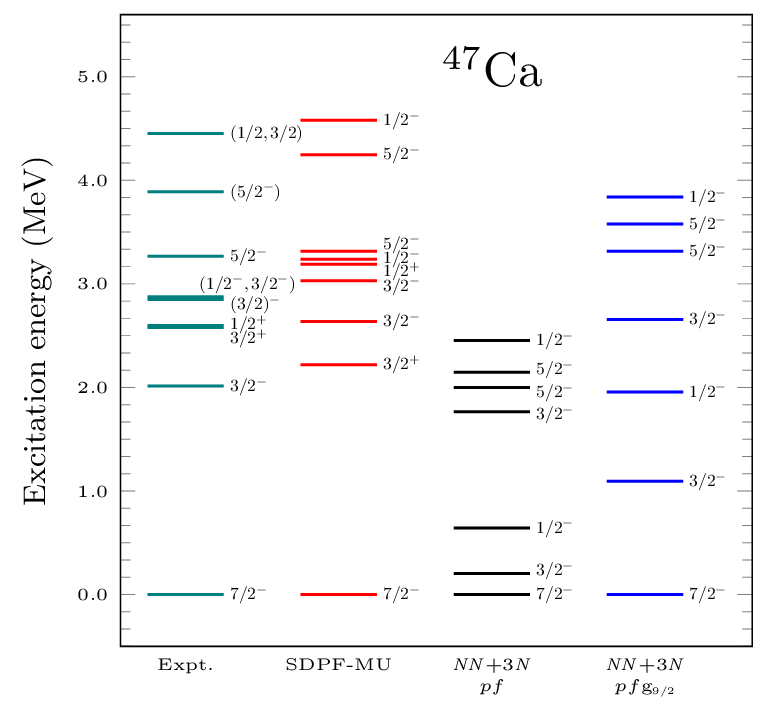}
	\caption{Comparison of excitation energies of low-lying excited states in $^{47}$Ca obtained from SDPF-MU interaction with previously obtained results in $pf$ and $pf$g$_{9/2}$ valence spaces using \textit{NN}+3\textit{N} forces \cite{PRC90}. Experimental data are also shown in first column for comparison.}
	\label{Comparison2}
\end{figure}

In Figs. \ref{Comparison1}-\ref{Comparison2}, we have shown a comparison of  the energy levels obtained from our present calculations with those of previous works, 
where the level scheme was obtained in $sd_{3/2}+f_{7/2}p_{3/2}$ \cite{46Ca}, $pf$ \cite{46Ca} and $pf$g$_{9/2}$ \cite{PRC90} valence spaces using \textit{NN}+3\textit{N} interactions. As can be seen from Fig. \ref{Comparison1}, although the lower excited states are well reproduced by previous calculation in $sd_{3/2}+f_{7/2}p_{3/2}$ model space \cite{46Ca}, the higher states, however, are  better reproduced by SDPF-MU interaction. For instance, our shell-model results with SDPF-MU interaction reported in this paper for $3^{-}$ state of $^{46}$Ca is in excellent agreement with the experimental value. 
 Negative parity states were not obtained below 5 MeV for  $pf$g$_{9/2}$ shell-model space.

 For $^{47}$Ca, previous studies were performed in $pf$ and $pf$g$_{9/2}$ valence spaces using \textit{NN}+3\textit{N} forces \cite{PRC90},  where only negative parity states were calculated. In the present work, we have calculated both positive and negative parity states. From Fig. \ref{Comparison2}, it is seen, for instance, that previous work with $pf$ and $pf$g$_{9/2}$ valence spaces give $\sim$ 1.8 MeV and $\sim$ 1 MeV energy difference between theory and experiment for $3/2^-_1$ state in $^{47}$Ca, respectively, while the difference is reduced to $\sim$ 600 keV in this work. This indicates that the $sdpf$-model space  is a suitable valence space for $^{47}$Ca. For other states also, we obtain improved results with SDPF-MU interaction as compared to the previous study.

We have analyzed configuration for low-lying positive and negative parity states of $^{46,47}$Ca corresponding to 1p-1h excitation.  
For natural parity states, there is no excitation from $sd$-shell to $pf$-shell. For $^{47}$Ca, $7/2^{-}_1$ and $3/2^{-}_1$ states are generated from the configurations  $\nu f_{7/2}^{-1}$ with probability of 94.41\%  and $\nu p_{3/2}^1$ with probability of 92.22\%, respectively. These states correspond to single particle states.
In  the case of unnatural parity states, either one proton or one neutron is excited into $pf$-shell. It is expected that for calcium isotopes, which have a proton shell closure at $Z = 20$, protons might not be excited from $sd$-shell to $pf$-shell. But for these states, we have seen that not only neutrons but also protons are excited from $sd$-shell to $pf$-shell. In fact, major contribution comes from proton excitation despite $Z =20$ is magic number. The $3^-_1$ state in $^{46}$Ca originated from the configuration $\pi(d_{5/2}^6s_{1/2}^1d_{3/2}^4f_{7/2}^1)$ $\otimes$ $\nu(d_{5/2}^6s_{1/2}^2d_{3/2}^4f_{7/2}^6)$, which has 22.45\%  contribution. Dominant configuration for  5/2$^+_{1}$ in $^{47}$Ca is   $\pi(d_{5/2}^6s_{1/2}^2d_{3/2}^3f_{7/2}^1)$ $\otimes$ $\nu(d_{5/2}^6s_{1/2}^2d_{3/2}^4f_{7/2}^7)$  with probability of 24.06\%.
So, protons are not treated as inert in $^{46,47}$Ca. This effect is also seen in the case of 2p-2h excitations. Even for natural parity states, protons are excited from $sd$-shell to $pf$-shell.

\section{Conclusion}
In the present work, we have performed a systematic nuclear shell-model calculation for recently available experimental data, measured with the GRIFFIN spectrometer at TRIUMF-ISAC facility from the
$\beta^-$ decay of $^{46}$K and $^{47}$K, to reproduce the level schemes of $^{46}$Ca and $^{47}$Ca.  We have also calculated low-lying energy spectra for the parent nuclei $^{46}$K and $^{47}$K.
In our shell-model calculations, the SDPF-MU interaction has been used to calculate spectroscopic properties of $^{46,47}$K and $^{46,47}$Ca. 
For the first time, we have calculated the $\log ft$ values corresponding to new data for the $\beta^-$ decay from $^{46,47}$K to $^{46,47}$Ca. 
Present shell-model study helps us to identify spin-parity of unconfirmed states 
for which the $\log ft$ values are reported in these experimental works. 
Based on calculated $\log ft$ values, we have predicted a state with 3.984 MeV to be either $3^-$ or $2^+$ with $\log ft$ value 6.009 and 9.080 for $^{46}$Ca.
A tentative state at 4.432 MeV is also assigned to be $3^-$ in $^{46}$Ca. In the case of $^{47}$Ca, we have also given spin confirmation to the tentative states at 2.850, 3.889 and 4.606 MeV. The states at 2.875, 3.951 and 4.453 MeV could be  $1/2^{-}$, $3/2^{-}$ and $1/2^{-}$ or $1/2^+$ in $^{47}$Ca, respectively. 
 The agreement of our calculated $\log ft$ values with the experimental data is rather limited.  
In some cases, the calculated $\log ft$ values are close to the experimental ones; the difference in $\log ft$ is within 0.8 for about half of the states measured in $^{46}$Ca and $^{47}$Ca. The difference, however, amounts to be 1.1-2.7 for other cases. 
Present nuclear shell-model study have added more information to the recent experimental works \cite{46Ca,47Ca}.

\section*{Acknowledgement}
P.C. acknowledges financial support from

MHRD (Government of India) for her Ph.D. thesis work.
P.C.S. acknowledges a research grant from the Science and Engineering
Research Board (SERB) of India, CRG/2019/000556. T.
S. acknowledges JSPS KAKENHI Grant JP19K03855. We
would like to thank J.L. Pore for useful discussions during
this work.

\bibliographystyle{utphys}
  \bibliography{references}

\begin{thebibliography}{44}
\expandafter\ifx\csname natexlab\endcsname\relax\def\natexlab#1{#1}\fi
\expandafter\ifx\csname bibnamefont\endcsname\relax
  \def\bibnamefont#1{#1}\fi
\expandafter\ifx\csname bibfnamefont\endcsname\relax
  \def\bibfnamefont#1{#1}\fi
\expandafter\ifx\csname citenamefont\endcsname\relax
  \def\citenamefont#1{#1}\fi
\expandafter\ifx\csname url\endcsname\relax
  \def\url#1{\texttt{#1}}\fi
\expandafter\ifx\csname urlprefix\endcsname\relax\def\urlprefix{URL }\fi
\providecommand{\bibinfo}[2]{#2}
\providecommand{\eprint}[2][]{\url{#2}}

\bibitem{A.Gade}
A. Gade, R. V. F. Janssens, D. Bazin, R. Broda, B. A. Brown, C. M. Campbell, M. P. Carpenter, J. M. Cook, A. N. Deacon, D.-C. Dinca \textit{et al.}, ``Cross-shell excitation in two-proton knockout: Structure of $^{52}$Ca '',
\href{https://journals.aps.org/prc/pdf/10.1103/PhysRevC.74.021302}
{Phys. Rev. C {\bf 74}, 021302 (2006)}.

\bibitem{F.Wienholtz}
F. Wienholtz, D. Beck, K. Blaum, Ch. Borgmann, M. Breitenfeldt, R. B. Cakirli, S. George, F. Herfurth, J. D. Holt, M. Kowalska \textit{et al.}, ``Masses of the exotic calcium isotopes pin down nuclear forces '',
\href{https://www.nature.com/articles/nature12226} 
{Nature {\bf 498}, 346-349 (2013)}.

\bibitem{D.Steppenbeck}
D. Steppenbeck, S. Takeuchi, N.Aoi, P. Doornenbal, M. Matsushita, H. Wang, H. Baba, N. Fukuda, S.Go, M. Honma  \textit{et al.}, ``Evidence for a new nuclear `magic' from the level structure of $^{54}$Ca'', \href{https://www.nature.com/articles/nature12522} 
{Nature {\bf 502}, 207-210 (2013)}.


\bibitem{D.SteppenbeckPRL}
D. Steppenbeck, S. Takeuchi, N. Aoi, P. Doornenbal, M. Matsushita, H. Wang, Y. Utsuno, H. Baba, S. Go, J. Lee \textit{et al.}, ``Low-lying structure of $^{50}$Ar and the $N = 32$ subshell closure'', \href{https://journals.aps.org/prl/abstract/10.1103/PhysRevLett.114.252501} 
{Phys. Rev. Lett. {\bf 114}, 252501 (2015)}.

\bibitem{M.Rosenbusch}
M. Rosenbusch, P. Ascher, D. Atanasov, C. Barbieri, D. Beck, K. Blaum, Ch. Borgmann, M. Breitenfeldt, R. B. Cakirli, A. Cipollone \textit{et al.}, ``Probing the $N = 32$ shell closure below the magic proton number $N = 20$: Mass measurements of the exotic isotopes $^{52,53}$K''
\href{https://journals.aps.org/prl/abstract/10.1103/PhysRevLett.114.202501}
{Phys. Rev. Lett. {\bf 114}, 202501 (2015)}.

\bibitem{H.N.Liu}
H. N. Liu, A. Obertelli, P. Doornenbal, C. A. Bertulani, G. Hagen, J. D. Holt, G. R. Jansen, T. D. Morris, A. Schwenk, R. Stroberg, \textit{et al.}, ``How robust in the $N = 34$ subshell closure? First spectroscopy of $^{52}$Ar''
\href{https://journals.aps.org/prl/abstract/10.1103/PhysRevLett.122.072502}
{Phys. Rev. Lett. {\bf 122}, 072502 (2019)}.


\bibitem{A.Ozawa}
A. Ozawa, T. Kobayashi, T. Suzuki, K. Yoshida, and I. Tanihata, ``New magic number, $N = 16$, near the neutron drip line'',
\href{https://journals.aps.org/prl/abstract/10.1103/PhysRevLett.84.5493}
{Phys. Rev. Lett. {\bf 84}, 5493 (2000)}.

\bibitem{R.Kanungo}
R. Kanungo, C. Nociforo, A. Prochazka, T. Aumann, D. Boutin, D. Cortina-Gil, B. Davids, M. Diakaki, F. Farinon, H. Geissel \textit{et al.}, ``One-neutron removal measurement reveals $^{24}$O as a new doubly magic nucleus'',
\href{https://journals.aps.org/prl/abstract/10.1103/PhysRevLett.102.152501}
{Phys. Rev. Lett. {\bf 102}, 152501 (2009)}.

\bibitem{A.Navin}
A. Navin, D. W. Anthony, T. Aumann, T. Baumann, D. Bazin, Y. Blumenfeld, B. A. Brown, T. Glasmacher, P. G. Hansen, R. W. Ibbotson \textit{et al.}, ``Direct evidence for the breakdown of the $N = 8$ shell closure in $^{12}$Be'',
\href{https://journals.aps.org/prl/abstract/10.1103/PhysRevLett.85.266}
{Phys. Rev. Lett. {\bf 85}, 266 (2000)}.

\bibitem{T.Motobayashi}
T.Motobayashi, Y. Ikeda, Y. Ando, K. Ieki, M. Inoue, N. Iwasa, T. Kikuchi, M. Kurokawa, S. Moriya, S. Ogawa \textit{et al.}, ``Large deformation of the very neutron-rich nucleus $^{32}$Mg from intermediate-energy Coulomb excitation'',
\href{https://www.sciencedirect.com/science/article/pii/037026939500012A?via}
{Phys. Lett. B{\bf 346}, 9 (1995)}.

\bibitem{B.Bastin}
B. Bastin, S. Gre{\'v}y, D. Sohler, O. Sorlin, Zs. Dombr{\'a}di, N. L. Achouri, J. C. Ang{\'e}lique, F. Azaiez, D. Baiborodin, R. Borcea \textit{et al.}, ``Collapse of the $N = 28$ shell closure in $^{42}$Si'',
\href{https://journals.aps.org/prl/abstract/10.1103/PhysRevLett.99.022503}
{Phys. Rev. Lett. {\bf 99}, 022503 (2007)}.



\bibitem{jouni}
J. Suhonen, {\it From Nucleons to Nucleus: Concept of Microscopic Nuclear Theory},
 (Springer, Berlin 2007).


\bibitem{mika2017}
M. Haaranen, J. Kotila, and J. Suhonen,
``Spectrum-shape method and the next -to-leading-order terms of the $\beta$-decay shape factor'',
\href{https://link.aps.org/doi/10.1103/PhysRevC.95.024327}
{Phys. Rev. C {\bf {95}} 024327 (2017)}.

\bibitem{mst2006}
M. T. Mustonen, M. Aunola, and J. Suhonen,
``Theoretical description of the fourth-forbidden non-unique $\beta$ decays of $^{113}$Cd and $^{115}$In'',
\href{https://link.aps.org/doi/10.1103/PhysRevC.73.054301}
{Phys. Rev. C {\bf {73}} 054301 (2006)};






\bibitem{mika2016}
M. Haaranen, P. C. Srivastava, and J. Suhonen
``Forbidden non-unique $\ensuremath{\beta}$ decays and effective values of weak coupling constants'',
\href{https://link.aps.org/doi/10.1103/PhysRevC.93.034308}
{Phys. Rev. C {\bf {93}} 034308 (2016)}.


\bibitem{joel12017} 
J. Kostensalo, M. Haaranen, and Jouni Suhonen,
``Electron spectra in forbidden $\beta$ decays and the quenching of the weak axial-vector coupling constant $\text{g}_A$'',
\href{https://link.aps.org/doi/10.1103/PhysRevC.95.044313}
{Phys. Rev. C {\bf 95}, 044313 (2017).}

\bibitem{joel22017} 
J. Kostensalo and J. Suhonen,
``$\text{g}_A$-driven shapes of electron spectra of forbidden $\beta$ decays in the nuclear shell-model'',
\href{https://link.aps.org/doi/10.1103/PhysRevC.96.024317}
{Phys. Rev. C {\bf 96}, 024317 (2017).}


\bibitem{VikashK}
V. Kumar, P.C. Srivastava, and H. Li, 
``Nuclear $\beta^-$ decay half-lives for $fp$ and $fp$g shell nuclei'',
\href{https://iopscience.iop.org/article/10.1088/0954-3899/43/10/105104}
{Jour. Phys. G: Nucl. and Part. Phys. {\bf 43}, 105104 (2016)}.

\bibitem{Anil} 
A. Kumar, P. C. Srivastava, J. Kostensalo and J. Suhonen,
``Second-forbidden non-unique $\beta^{-}$
decays of $^{24}$Na and $^{36}$Cl assessed by the nuclear shell-model'',
\href{https://journals.aps.org/prc/abstract/10.1103/PhysRevC.101.064304}
{Phys. Rev. C {\bf 101}, 064304 (2020).}


\bibitem{W.W.Daehnick}
W. W. Daehnick and M. J. Spisak,
``A look at the $^{46}\mathrm{Ca}$ level structure with $^{46}$Ca($p,p^{'}$)$^{46}$Ca at $E_p = 16 $ MeV'',
\href{https://journals.aps.org/prc/abstract/10.1103/PhysRevC.7.2593}
{Phys. Rev C {\bf 7}, 2593 (1973)}.

\bibitem{D.C.Williams}
D. C. Williams, J. D. Knight, and W. T. Leland,
``Levels of $^{42}\mathrm{Ca}$ and $^{46}\mathrm{Ca}$ as observed in the $^{40}$Ca($t,p$) and $^{44}$Ca($t,p$) reactions'',
\href{https://journals.aps.org/pr/abstract/10.1103/PhysRev.164.1419}
{Phys. Rev. {\bf 164}, 1419 (1967)}.



\bibitem{G.M.Crawley}
G. M. Crawley, P. S. Miller, G. J. Igo, and J. Kulleck,
``High-resolution study of $^{48}\mathrm{Ca}$($p,t$)$^{46}\mathrm{Ca}$ at $E_{p}$ = 39 MeV'',
\href{https://journals.aps.org/prc/abstract/10.1103/PhysRevC.8.574}
{Phys. Rev. C {\bf 8}, 574 (1973)}.



\bibitem{M.E.Williams-Norton}
M. E. Williams-Norton and R. Abegg,
``Study of low-lying levels in $^{47}\mathrm{Ca}$ with the $^{48}\mathrm{Ca}$($d,t$)$^{47}\mathrm{Ca}$ reaction'',
\href{https://www.sciencedirect.com/science/article/abs/pii/037594747790330X?via%3Dihub}
{Nucl. Phys. A {\bf 291}, 429 (1977)}.

\bibitem{P.Martin}
P. Martin, M. Buenerd, Y. Dupont, and M. Chabre,
`Proton-deuteron reactions at 40 MeV on the calcium isotopes'',
\href{https://www.sciencedirect.com/science/article/abs/pii/0375947472900255?via%3Dihub}
{Nucl. Phys. A {\bf 185}, 465 (1972)}.


\bibitem{B.Parsa}
B. Parsa and G.E. Gorden,
``A new nucleide: 115 sec $^{46}\mathrm{K}$'',
\href{https://www.sciencedirect.com/science/article/pii/0031916366901806?via}
{Phys. Lett. {\bf 23}, 269 (1966)}.

\bibitem{P.Kunz}
P. Kunz, C. Andreoiu, P. Bricault, M. Dombsky, J. Lassen, A. Teigelh{\"o}fer, H. Heggen, and F. Wong,
``Nuclear and in-source laser spectroscopy with the ISAC yield station'',
\href{https://aip.scitation.org/doi/10.1063/1.4878718}
{Rev. Sci. Instrum. {\bf 85}, 053305 (2014)}.

\bibitem{E.K.Warburton}
E. K. Warburton, D. E. Alburger, and G. A. P. Engelbertink,
``Beta decay of $^{47}\mathrm{K}$, $^{50}\mathrm{Ca}$ and $^{50}\mathrm{Sc}$'',
\href{https://journals.aps.org/prc/abstract/10.1103/PhysRevC.2.1427}
{Phys. Rev. C {\bf 2}, 1427 (1970)}.

\bibitem{D.E.Alburger}
D. E. Alburger, E. K. Warburton, and B. A. Brown,
``Decay of $^{50}\mathrm{Sc}$, $^{50}\mathrm{Sc}^{m}$, $^{50}\mathrm{Ca}$ and $^{47}\mathrm{K}$'',
\href{https://journals.aps.org/prc/abstract/10.1103/PhysRevC.30.1005}
{Phys. Rev. C {\bf 30}, 1005 (1984)}.

\bibitem{T.Kuroyanagi}
T. Kuroyanagi, T. Tamura, K. Tanaka, and H. Morinaga,
``Potassium-47'',

\href{https://www.sciencedirect.com/science/article/abs/pii/0029558264902184?via%3Dihub}
{Nucl. Phy. {\bf 50}, 417 (1964)}.


\bibitem{46Ca}
J.L. Pore, C. Andreoiu, J. K. Smith, A. D. MacLean, A. Chester, J. D. Holt, G. C. Ball, P. C. Bender, V. Bildstein, R. Braid \textit{et al.},
``Detailed spectroscopy of $^{46}\mathrm{Ca}$: A study of the ${\ensuremath{\beta}}^{\ensuremath{-}}$ decay of $^{46}\mathrm{K}$'',
\href{https://link.aps.org/doi/10.1103/PhysRevC.100.054327}
{Phys. Rev. C {\bf 100}, 054327 (2019)}.

\bibitem{47Ca}
J.K. Smith, A. B. Garnsworthy, J. L. Pore, C. Andreoiu, A. D. MacLean, A. Chester, Z. Beadle, G. C. Ball, P. C. Bender, V. Bildstein \textit{et al.},
``Spectroscopic study of $^{47}\mathrm{Ca}$ from the ${\ensuremath{\beta}}^{\ensuremath{-}}$ decay of $^{47}\mathrm{K}$'',
\href{https://link.aps.org/doi/10.1103/PhysRevC.102.054314}
{Phys. Rev. C {\bf 102}, 054314 (2020)}.


\bibitem{nocore_review} B. R. Barrett, P. Navr\'atil and J. P. Vary, ``\textit{Ab initio} no core shell-model'', 
\href{https://doi.org/10.1016/j.ppnp.2012.10.003}
{Prog. Part. Nucl. Phys. {\bf 69}, 131 (2013).}

\bibitem{P.Choudhary}
P.Choudhary, P. C. Srivastava and P. Navr{\'a}til, ``\textit{Ab initio} no-core shell-model study of $^{10-14}$B isotopes with realistic \textit{NN} interactions'', 
\href{https://journals.aps.org/prc/abstract/10.1103/PhysRevC.102.044309}
{Phys. Rev. C {\bf 102}, 044309 (2020).}

\bibitem{arch2} A. Saxena and P. C. Srivastava, ``\textit{Ab initio} no-core shell-model study of $^{18-23}$O and $^{18-24}$F isotopes'', 
\href{https://doi.org/10.1088/1361-6471/ab6f1d}
{J Phys. G: Nucl. Part. Phys. {\bf 47}, 055113 (2020).}

\bibitem{G.HagenPRL}G. Hagen, G. R. Jansen, and T. Papenbrock, ``Structure of $^{78}$Ni from first-principles computations'', 
\href{https://journals.aps.org/prl/abstract/10.1103/PhysRevLett.117.172501}
{Phys. Rev. Lett. {\bf 117}, 172501 (2016).}

\bibitem{T.D.Morris}T. D. Morris, J. Simonis, S. R. Stroberg, C. Stumpf, G. Hagen, J. D. Holt, G. R. Jansen, T. Papenbrock, R. Roth, and A.
Schwenk, ``Structure of the lightest  Tin isotopes'', 
\href{https://journals.aps.org/prl/abstract/10.1103/PhysRevLett.120.152503}
{Phys. Rev. Lett. {\bf 120}, 152503 (2018).}

\bibitem{Holtdrip}
S. R. Stroberg, J. D. Holt, A. Schwenk, and J. Simonis, ``\textit{Ab initio} limits of atomic nuclei'',
\href{https://doi.org/10.1103/PhysRevLett.126.022501}
{Phys. Rev. Lett. {\bf 126}, 022501 (2021).}

\bibitem{Anil2020}
A. Kumar, P.C. Srivastava and T. Suzuki, ``Shell model results for nuclear $\beta^-$-decay properties of sd-shell nuclei'',  
\href{https://doi.org/10.1093/ptep/ptaa012}
{Prog. Theo. Expt. Phys. {\bf {2020}},  033D01 (2020)}.





\bibitem{G.Hagen}
G. Hagen, M. H.-Jensen, G. R. Jansen, R. Machleidt, and T. Papenbrock,
``Evolution of shell structure in neutron-rich calcium isotopes''
\href{https://doi.org/10.1103/PhysRevLett.109.032502}
{Phys. Rev. C{\bf 109}, 032502 (2012)}.


\bibitem{RMP}E. Epelbaum, H.-W. Hammer and U.-G. Mei\ss ner, ``Modern theory of nuclear forces'', 
\href{https://doi.org/10.1103/RevModPhys.81.1773}
{Rev. Mod. Phys. {\bf 81}, 1773 (2009).}

\bibitem{EFT1}S. Weinberg, ``Phenomenological Lagrangians'', 
\href{https://doi.org/10.1016/0378-4371(79)90223-1}
{Physica A {\bf 96}, 327 (1979).}



\bibitem{EFT2}S. Weinberg, ``Nuclear forces from chiral lagrangians'', 
\href{https://doi.org/10.1016/0370-2693(90)90938-3}
{Phys. Lett. B {\bf 251}, 288 (1990).}

\bibitem{EFT3}S. Weinberg, ``Effective chiral lagrangians for nucleon-pion interactions and nuclear forces'', 
\href{https://doi.org/10.1016/0550-3213(91)90231-L}
{ Nucl. Phys. B {\bf 363}, 3 (1991).}



\bibitem{J.D.Holt}
J. D. Holt, T. Otsuka, A. Schwenk and T. Suzuki,
"Three-body forces and shell structure in calcium
isotopes"
\href{https://doi.org/10.1088/0954-3899/39/8/085111}
{J. Phys. G: Nucl. Part. Phys. {\bf 39}, 085111 (2012)}.


\bibitem{PRC90}
J. D. Holt, J. Men\'endez, J. Simonis, and A. Schwenk,
``Three-nucleon forces and spectroscopy of neutron-rich calcium isotopes''
\href{https://doi.org/10.1103/PhysRevC.90.024312}
{Phys. Rev. C {\bf 90}, 024312 (2014)}.



\bibitem{M.Honma}
M. Honma, T. Otsuka, B. A. Brown and T. Mizusaki,
``Shell-model description of neutron-rich $pf$-shell nuclei with a new effective interaction''
\href{https://link.springer.com/article/10.1140/epjad/i2005-06-032-2}
{Eur. Phys. J A {\bf 25}, 499 (2005)}.

\bibitem{A.Poves}
 A. Poves, J. S{\'a}nchez-Solano, E. Caurier and F. Nowacki,
``Shell-model study of the isobaric chains $A=50$, $A=51$ and $A=52$''
\href{https://www.sciencedirect.com/science/article/pii/S0375947401009678}
{Nucl. Phys. A {\bf 694}, 157 (2001)}.

\bibitem{S.R.Stroberg}
S. R. Stroberg, A. Calci, H. Hergert, J. D. Holt, S. K. Bogner, R. Roth,  and A. Schwenk,
``Nucleus-Dependent valence-space approach to nuclear structure''
\href{https://doi.org/10.1103/PhysRevLett.118.032502}
{Phys. Rev. Lett. {\bf 118}, 032502 (2017)}.

\bibitem{V.Soma}
V. Som{\'a}, P. Navr{\'a}til, F. Raimondi, C. Barbieri, and T. Duguet,
``Novel chiral Hamiltonian and observables in light and medium-mass nuclei'',
\href{https://doi.org/10.1103/PhysRevC.101.014318}
{Phys. Rev. C {\bf 101}, 014318 (2020)}.



\bibitem{T.Togashi}
T. Togashi, N. Shimizu, Y. Utsuno, T. Otsuka, and M. Honma
"Large-scale shell-model calculations for unnatural-parity high-spin states in neutron-rich Cr and Fe isotopes
\href{https://journals.aps.org/prc/abstract/10.1103/PhysRevC.91.024320}
{Phys. Rev. C {\bf 91}, 024320 (2015)}.

\bibitem{B.Bhoy}
B. Bhoy, P.C. Srivastava and K. Kaneko,
"Shell-model results for $^{47-58}$Ca isotopes in the $fp$, $fpg_{9/2}$ and $fpg_{9/2}d_{5/2}$-model spaces "
\href{https://doi.org/10.1088/1361-6471/ab80d4}
{J. Phys. G: Nucl. Part. Phys. {\bf 47}, 065105 (2020)}.


\bibitem{Nature}
R. F. Garcia Ruiz, M. L. Bissell, K. Blaum, A. Ekstr{\"o}m, N. Fr{\"o}mmgen, G. Hagen, M. Hammen, K. Hebeler, J. D. Holt, G. R. Jansen \textit{et al.}
``Unexpectedly large charge radii of neutron-rich calcium isotopes'',
\href{https://www.nature.com/articles/nphys3645}
{Nature Physics {\bf 12}, 594-598 (2016)}.

\bibitem{L.Coraggio}
L. Coraggio, G. De Gregorio, A. Gargano, N. Itaco, T. Fukui, Y. Z. Ma, and F. R. Xu,
``Shell-model study of calcium isotopes towards their drip line'',
\href{https://doi.org/10.1103/PhysRevC.102.054326}
{Phys. Rev. C {\bf 102}, 054326 (2020)}.


\bibitem{SDPFMU}
Y. Utsuno, T. Otsuka, B. A. Brown, M. Honma, T. Mizusaki, and N. Shimiz
``Shape transitions in exotic Si and S isotopes and tensor-force-driven Jahn-Teller effect'',
\href{https://journals.aps.org/prc/abstract/10.1103/PhysRevC.86.051301}
{Phys. Rev. C {\bf 86}, 051301(R) (2012)}.

 \bibitem{behrens1982}
H.~Behrens and W.~B$\ddot{\text{u}}$hring,  {\it Electron Radial Wave Functions and Nuclear Beta-Decay} (Clarendon, Oxford, 1982).


\bibitem{hfs1966}
H. F. Schopper, {\it Weak Interaction and Nuclear Beta Decay} (North-Holland, Amsterdam, 1966).

  \bibitem{Patrignani}
C. Patrignani and Particle Data Group,
``Review of Particle Physics'',
\href{https://doi.org/10.1088/1674-1137/40/10/100001}
{Chinese Phys. C {\bf 40}, 100001 (2016).}



\bibitem{nushellx}
B. A. Brown and W. D. M. Rae, The shell-model code
NuShellX@MSU, \href{https://doi.org/10.1016/j.nds.2014.07.022}
{Nucl. Data Sheets 120, 115 (2014).}

\bibitem{enhancement}
E. K. Warburton, 
``In-medium and core-polarization effects in $^{50}$K($0^-$) $\rightarrow$ $^{50}$Ca($0^+$)'',
\href{https://journals.aps.org/prc/pdf/10.1103/PhysRevC.44.1024}
{Phys. Rev. C {\bf 44}, 233 (1991)}.

\bibitem{PRC85}
T. Suzuki, T. Yoshida, T. Kajino, and T. Otsuka, 
``$\beta$ decays of isotones with neutron magic number of $N$=126 and $r$-process nucleosynthesis'',
\href{https://journals.aps.org/prc/abstract/10.1103/PhysRevC.85.015802}
{Phys. Rev. C {\bf 85}, 015802 (2012)}.


\bibitem{Towner}  I. S. Towner, 
``Enhancement in axial-charge matrix elements from meson-exchange currents'',
\href{https://doi.org/10.1016/0375-9474(92)90261-H}
{Nucl. Phys. A 542, 631 (1992).}


\bibitem{WTB} E. K. Warburton, I. S. Towner and B. A. Brown,
``First-forbidden \ensuremath{\beta} decay: Meson-exchange enhancement of the axial charge at A\ensuremath{\sim}16'', 
\href{https://link.aps.org/doi/10.1103/PhysRevC.49.824}
{Phys. Rev. C 49, 824 (1994).}


\bibitem{meson}
P. Baumann, M. Bounajma, F. Didierjean, A. Huck, A. Knipper, M. Ramdhane, and G. Waltera, 
``Meson-exchange enhancement in first-forbidden $\beta$ transitions: The case of $^{50}$K and $^{38}$Ca'',
\href{https://journals.aps.org/prc/abstract/10.1103/PhysRevC.58.1970}
{Phys. Rev. C {\bf 58}, 1970 (1998)}.

\bibitem{NNDC}Data extracted using the NNDC World Wide Web site from the ENSDF, 
 \href{https://www.nndc.bnl.gov/ensdf/.}
 { https://www.nndc.bnl.gov/ensdf/.}	
 
\bibitem{Qandmag}IAEA, 
 \href{https://www-nds.iaea.org/nuclearmoments/.}
 {https://www-nds.iaea.org/nuclearmoments/.}
 

\bibitem{Honma}
M. Honma, T. Otsuka, B. A.Brown, and T. Mizusaki, 
``New effective interaction for $pf$-shell nuclei and its applications for the stability of the $N=Z$ = 28 closed core'',
\href{https://journals.aps.org/prc/abstract/10.1103/PhysRevC.69.034335}
{Phys. Rev. C {\bf 69}, 034335 (2004)}.

\bibitem{Richter}
W. A. Richter, S. Mkhize, and B.A. Brown, 
``sd-shell observables for the USDA and USDB Hamiltonians'',
\href{https://journals.aps.org/prc/abstract/10.1103/PhysRevC.78.064302}
{Phys. Rev. C {\bf 78}, 064302 (2008).}

\end{thebibliography}

\end{document}